\documentclass[aps,showpacs,amsmath,amssymb,superscriptaddress]{revtex4}

\usepackage{graphicx}
\usepackage{amssymb,amsmath}
\usepackage{dcolumn}
\usepackage{bm}
\begin{document}

\title{Influence of the mirrors on the strong coupling regime in planar GaN microcavities}

\author{F. R\'everet}
\affiliation{LASMEA, UMR CNRS-Universit\'e Blaise Pascal 6602, 24 Avenue des Landais, 63177 Aubi\`ere Cedex France}
\author{P. Disseix}
\affiliation{LASMEA, UMR CNRS-Universit\'e Blaise Pascal 6602, 24 Avenue des Landais, 63177 Aubi\`ere Cedex France}
\author{J. Leymarie}
\affiliation{LASMEA, UMR CNRS-Universit\'e Blaise Pascal 6602, 24 Avenue des Landais, 63177 Aubi\`ere Cedex France}
\author{A. Vasson}
\affiliation{LASMEA, UMR CNRS-Universit\'e Blaise Pascal 6602, 24 Avenue des Landais, 63177 Aubi\`ere Cedex France}
\author{F. Semond}
\affiliation{CRHEA-CNRS, Rue Bernard Gregory, Parc Sophia Antipolis, Valbonne, France}
\author{M. Leroux}
\affiliation{CRHEA-CNRS, Rue Bernard Gregory, Parc Sophia Antipolis, Valbonne, France}
\author{J. Massies}
\affiliation{CRHEA-CNRS, Rue Bernard Gregory, Parc Sophia Antipolis, Valbonne, France}

\begin{abstract}
The optical properties of bulk $\lambda/2$ GaN microcavities working in the strong light-matter coupling regime are investigated using angle-dependent reflectivity and photoluminescence at 5 K and 300 K. The structures have an Al$_{0.2}$Ga$_{0.8}$N/AlN distributed Bragg reflector as the bottom mirror and either an aluminium mirror or a dielectric Bragg mirror as the top one. First, the influence of the number of pairs of the bottom mirror on the Rabi splitting is studied. The increase of the mirror penetration depth is correlated with a reduction of the Rabi splitting. Second, the emission of the lower polariton branch is observed at low temperature in a microcavity containing two Bragg mirrors and exibiting a quality factor of 190. Our simulations using the transfer-matrix formalism, taking into account the real structure of the samples investigated are in good agreement with experimental results.
\end{abstract}

\pacs{78.66.Fd, 71.35.-y, 78.20.Ci}

\maketitle

\section{Introduction}
\label{sec:Introduction}

Since its first observation, the strong light-matter coupling regime (SCR) in semiconductor microcavities has been widely investigated, mainly in classical III-V and II-VI compounds \cite{Weisbuch, R.Houdre2005,M.S.Skolnick1998,PhysRevLett.81.3920}. This regime appears when photons and excitons having the same energy and momentum have time to strongly interact; a quasi-particle, called cavity-polariton, results in the mixing of these photon and exciton states. 
Non-linear effects and, more rencently, Bose-Einstein condensation have been observed in II-VI microcavities at low temperature \cite{PhysRevLett.81.3920,Kasprzak}.
The need to maintain the SCR at room temperature in order to realize the polariton laser has directed the interest towards semiconductors with large exciton binding energy and large oscillator strength as III-nitrides \cite{Imamoglu,G.Malpuech2002}. 
Since the first observation of the SCR at low temperature in a bulk GaN microcavity grown on silicon \cite{Vincent2003a}, growth techniques and design of mirrors have considerably improved, which has led to the SCR observation and polariton emission in GaN planar microcavities at room temperature \cite{F.Semond2005,I.R.Sellers2006,Gurioli,G.Christmann2006}.
Two kinds of structure have been used as bottom mirrors on sapphire substrate : a lattice-matched Al$_{0.85}$In$_{0.15}$N/Al$_{0.2}$Ga$_{0.85}$N distributed Bragg mirror (DBR) \cite{butte033315,Baumberg} or an Al$_{0.2}$Ga$_{0.8}$N/Al$_{0.6}$Ga$_{0.4}$N DBR \cite{alyamani:093110}.
Another possibility is to grow an Al$_{0.2}$Ga$_{0.8}$N/AlN DBR directly on a silicon substrate; the high index constrast between AlN and Si yields a high reflectivity with a DBR containing a lower AlGaN/AlN pair number \cite{Sellers}.
In such microcavity structures, the use of metallic, semiconductor or dielectric mirrors to confine photons in the active layer has a crucial influence on the finesse of the optical mode (or quality factor of the microcavity) and hence on the Rabi splitting.
In this work, we study the influence of these dielectric and/or metallic mirrors on the optical properties of bulk GaN microcavities. The polaritonic dispersion of various geometry microcavity structures is investigated through angle-resolved reflectivity and photoluminescence experiments at both low and room temperatures, and transfer matrix simulations are used to interpret the experimental data.
After a description of the samples studied, of the experimental setup and of the modelling of the reflectivity spectra in section 2, the influence of the bottom mirror reflectivity on the strong coupling regime is analysed in section 3. In section 4, the effect of increasing the cavity quality factor on their optical properties, in particular luminescence, will be presented.

\section{Samples description, experimental setup and simulation of reflectivity spectra.}
\label{sec:SampleDescriptionAndExperimentalSetup}

Four bulk GaN based microcavities (labelled A to D) grown by molecular beam epitaxy on (111) Si substrates have been studied. The thickness of the GaN active layer is $\lambda/2$ ($\lambda$ is the resonance wavelength in the cavity) and the cavity resonance \hbox{energy} is close to the excitonic one (3.45~eV or $\lambda_{0}$=359~nm). The bottom mirror is a distributed Bragg reflector (DBR) consisting of Al$_{0.2}$Ga$_{0.8}$N/AlN pairs directly grown on the substrate and starting with AlN. The number of $\lambda/2$ thick pairs of this bottom mirror is respectively equal to seven, ten and fifteen for samples A, B and C respectively. After the growth of the $\lambda/2$ bulk GaN cavity, a 10~nm thick semitransparent aluminium top mirror is evaporated.
The fourth sample (D) is the same as sample B but the aluminium mirror is replaced by a dielectric mirror containing eight $\lambda/2$ pairs of SiN/SiO$_{2}$. The substrate rotation was stopped during the epitaxy of the GaN active layer in order to create a thickness gradient, which we used to adjust the optical mode of the cavity for our experiments. The studied microcavities were designed to be slightly negatively detuned under normal incidence at the center of the wafer in order to observe the cavity resonance at both low and room temperature by tuning the incident angle.\\


Angle resolved reflectivity experiments were performed at 5 K and at room temperature. The sample is placed inside a rotating helium cryostat with quartz cylindrical windows. The light source is an halogen lamp mounted together with a linear polarizer on a mobile rail. The whole system allows polarization-dependent measurements with incident angle varying from $2^{\circ}$ and up to $85^{\circ}$.
A 64 cm focal length monochromator and a GaAs photomultiplier with lock-in amplification are used for the detection. The same set-up is used for angle-resolved photoluminescence experiments using an He-Cd laser (325~nm) with an angular resolution of about $1^{\circ}$.\\

The simulations of the reflectivity spectra were carried out within the transfer matrix formalism \cite{Azzam} using a complex refractive index \cite{Vincent2003}. The GaN excitonic transitions are modeled by inhomogeneously broadened Lorentz oscillators. The relative dielectric function is given by the following expression:

\begin{align}
\epsilon_{r}(E)=\epsilon_{b}+\sum_{j=A,B}\int\frac{1}{\sqrt{\pi}\Delta_{j}}~\frac{B_{j}}{x^{2}-E^{2}+i\hbar\Gamma_{j} E}~e^{\frac{-(x-E_{0j})^{2}}{\Delta^{2}_{j}}}dx
\end{align}

\vspace{1cm}

\noindent
where $\epsilon_{b}$ is the background dielectric constant. $B_{j}$, $E_{0j}$, $\Gamma_{j}$ and $\Delta_{j}$ are respectively the oscillator strength, the resonance energy, the homogeneous and the inhomogeneous broadening parameters of the $j$ excitonic level. Only two exciton states are included in the calculations since  a maximum of two are observed in this work. Should the active GaN layer be under compressive (0001) biaxial stress (27~kbar, due to the growth on Al$_{0.2}$Ga$_{0.8}$N/AlN DBRs) the oscillator strength of C exciton would be weak for near normal internal incidence \cite{B.Gil1997,O.Aoude2004}. Note that in order to account for the strain gradient along the growth axis, the active layer is divided in five layers of constant strain, the highest strain value corresponding to the bottom layer \cite{F.Semond2005}.

Besides these simulations of reflectivity spectra, the dispersion of the polariton branches was also calculated by using a [3x3] matrix formalism, where the matrix is given by \cite{I.R.Sellers2006}:

\vspace{0.5cm}

\begin{align}
\begin{pmatrix}
E_{ph}-j\Gamma_{ph}&V_{A}&V_{B}\\
V_{A}&E_{A}-j\Gamma_{A}&0 \\
V_{B}&0&E_{B}-j\Gamma_{B}
\end{pmatrix}
\end{align}

\vspace{1cm}

\noindent
$E_{ph}$, $E_{A}$, $E_{B}$, $\Gamma_{ph}$, $\Gamma_{A}$ and $\Gamma_{B}$ are respectively the energies of the photonic mode and the uncoupled A and B excitons and their associated homogeneous linewidths. $V_{A}$ and $V_{B}$ are the interaction terms responsible for the exciton-photon coupling.\\

\section{Influence of the bottom mirror reflectivity on the Rabi splitting}
\label{sec:Influenceofthebottommirror}

\begin{figure}[!ht]
		\centering
        \includegraphics[scale=0.83]{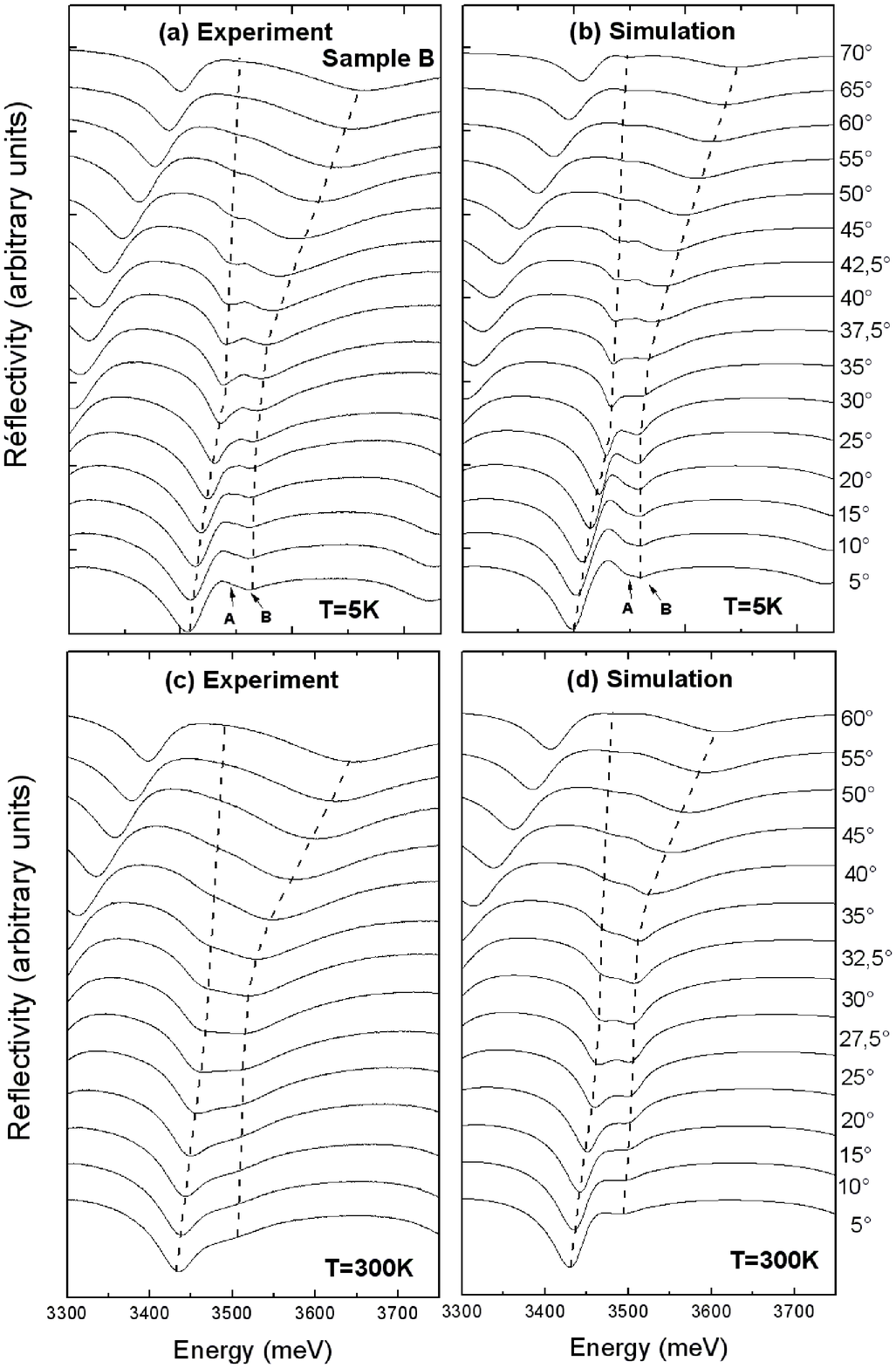}        
				\caption{}
        \label{fig:SampleB}
\end{figure}

Figure \ref{fig:SampleB}(a) displays the evolution with incident angle of the reflectivity spectra of sample B (ten pairs DBR) at 5 K and for the transverse electric (TE) polarization. Similar results are recorded for TM polarization (not shown).
For an incident angle of $5^{\circ}$, the optical mode with a quality factor Q (=$\lambda/\Delta\lambda$) of 60 is observed at $3459\pm3$~meV and the A and B excitonic dips at $3519\pm3$~meV and $3538\pm3$~meV respectively. 
The A-B splitting deduced from these spectra ($\sim19$~meV) is larger than the value expected (11-13~meV) for biaxially compressed GaN \cite{B.Gil1997,O.Aoudetobepublished}, suggesting that the stress on the active layer is not purely biaxial. 
With increasing incident angle, the optical mode blue-shifts and its anticrossing with the A and B excitonic modes is clearly observed at $35^{\circ}$ with a Rabi splitting $\Omega_{Rabi}$ of $49\pm2$~meV. Note that in this work, we define the Rabi splitting as the minimum difference between the energies of the lower and upper polariton branches, irrespective of whether the A and B excitonic levels are resolved (at 5 K) or not (at 300 K), i.e. whether the medium polariton branch is observed or not.\\
The simulations of the 5 K reflectivity spectra of sample B are displayed on figure~\ref{fig:SampleB}(b). The evolution of the photonic and the excitonic modes as function of the incident angle, as well as their anticrossings are excellently reproduced. It is worth noting that our modelling work accounts also for the evolution of the low-energy edge of the Bragg mirror, observed from $35^{\circ}$.
The A and B exciton energy separation used in the simulation is $18\pm1$~meV. For both excitons, the fitting oscillator strength value is $50000\pm5000$~meV$^{2}$, in good agreement with recent measurements \cite{O.Aoude2004}, and the homogeneous and inhomogeneous broadening parameters are respectively equal to $\gamma=0.1$~meV and $\Delta=12\pm2$~meV.
As previously mentioned, the introduction of a strain gradient along the growth axis gives a better agreement with the experimental results, particulary near the resonance. The GaN active layer has been divided in five discrete layers with an excitonic energy shift of 3 meV from one layer to the next. This results in a slight diminution of the excitonic linewidth from $2\sqrt{\ln2}\Delta$=25 to 20~meV in each slab.\\

\begin{figure}[!ht]
		\centering
        \includegraphics[scale=0.6]{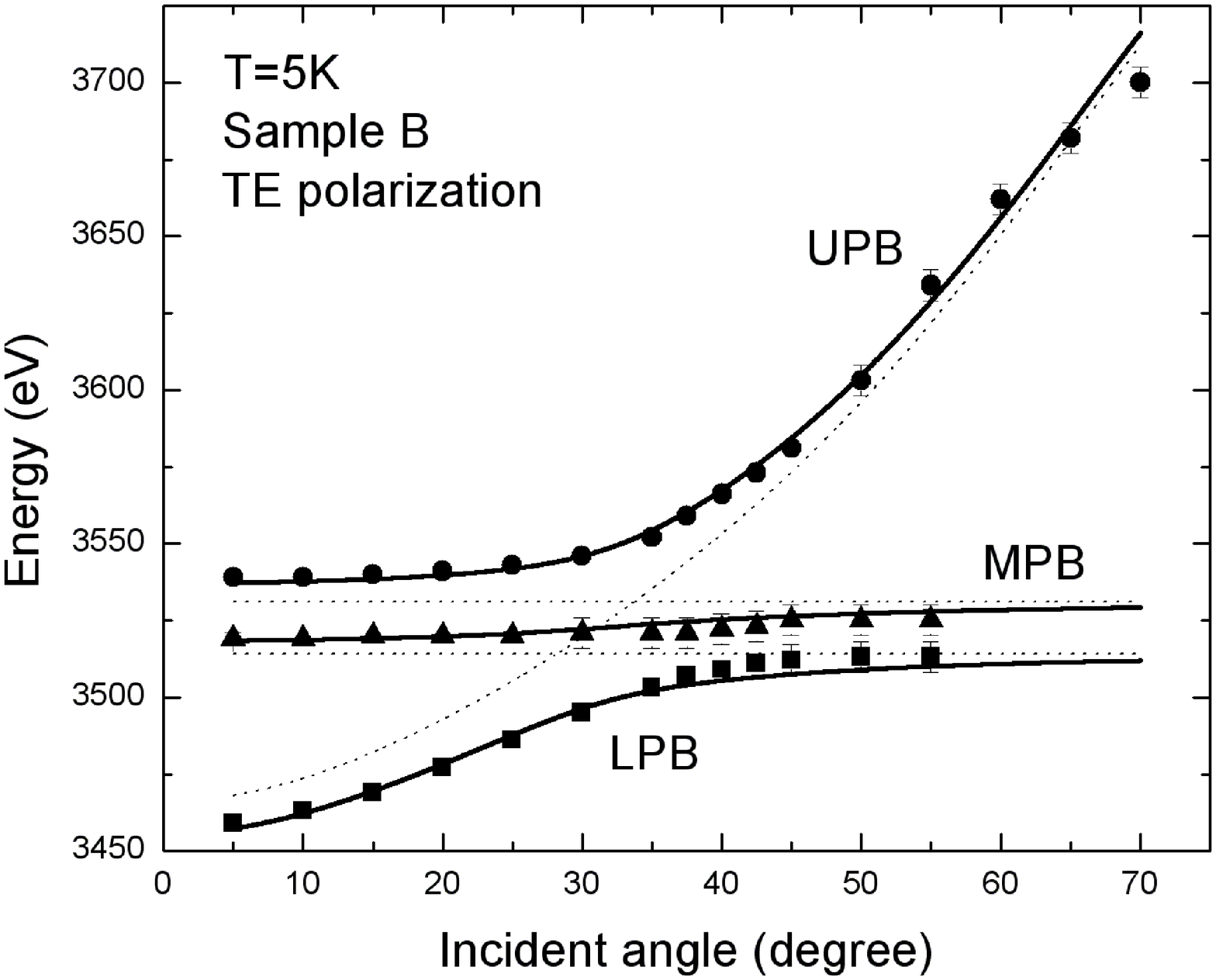}
				\caption{}
        \label{A808A_Modèle_3x3}
\end{figure}

The experimentally determined polariton energies at 5 K as function of incident angle are reported on figure~\ref{A808A_Modèle_3x3}.
The solid lines represent the results of the [3x3] matrix diagonalization (equation 2) with the following fitting parameter values \hbox{$E_{ph}=3465$~meV} at angle $\theta=0^{\circ}$, $E_{A}=3517$~meV, $E_{B}=3535$~meV, $\Gamma_{ph}=63$~meV and $\Gamma_{A}=\Gamma_{B}=0.1$~meV. The linewidth of the photonic mode was deduced from the $5^{\circ}$ \hbox{reflectivity} spectrum. The interaction potentials deduced from the fit are $V_{A}=V_{B}=23$~meV. The uncoupled A and B excitons together with the photonic mode are also reported as dotted lines. The evolution of the latter with the incident angle is given by \cite{M.S.Skolnick1998}:

\begin{align}
	E(\theta)=E_{ph}\left(1-\frac{\sin^2\theta}{n^{2}_{eff}}\right)^{-1/2}
\end{align}

\vspace{1cm}
\noindent
where $\theta$ and $n_{eff}$ are respectively the incidence angle and the effective index of the GaN active layer.
The index $n_{eff}$ is adjusted in order to obtain an uncoupled photonic dispersion in agreement with the experimental data away from the strong coupling region.
Results similar to those shown for sample B and for both polarizations were obtained regarding the angle dependence of the 5 K reflectivity spectra of samples A, C and D.

\begin{figure}[!ht]
		\centering
        \includegraphics[scale=0.8]{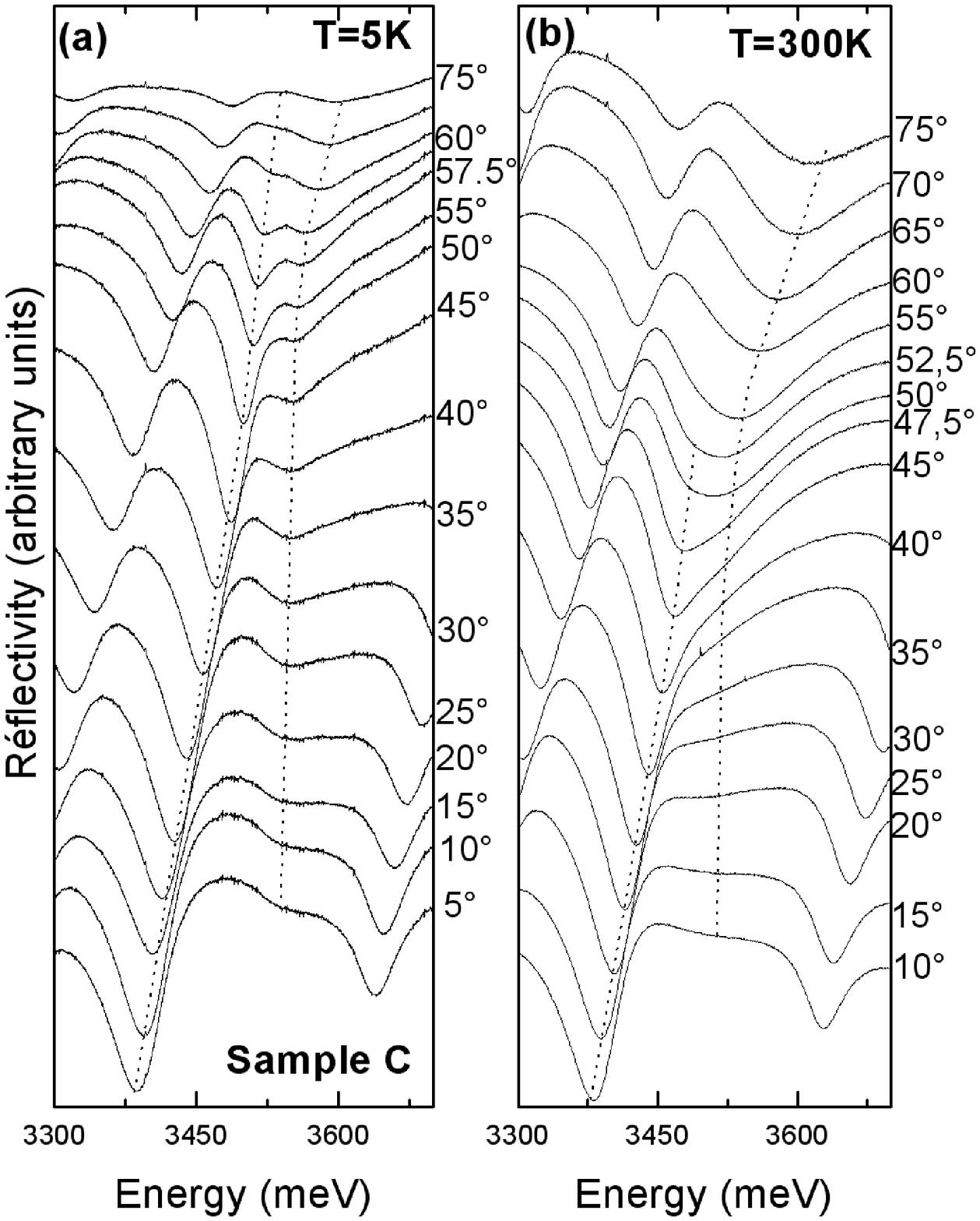}        
				\caption{}
        \label{fig:R_A771}
\end{figure}

\begin{figure}[!ht]
		\centering
        \includegraphics[scale=0.8]{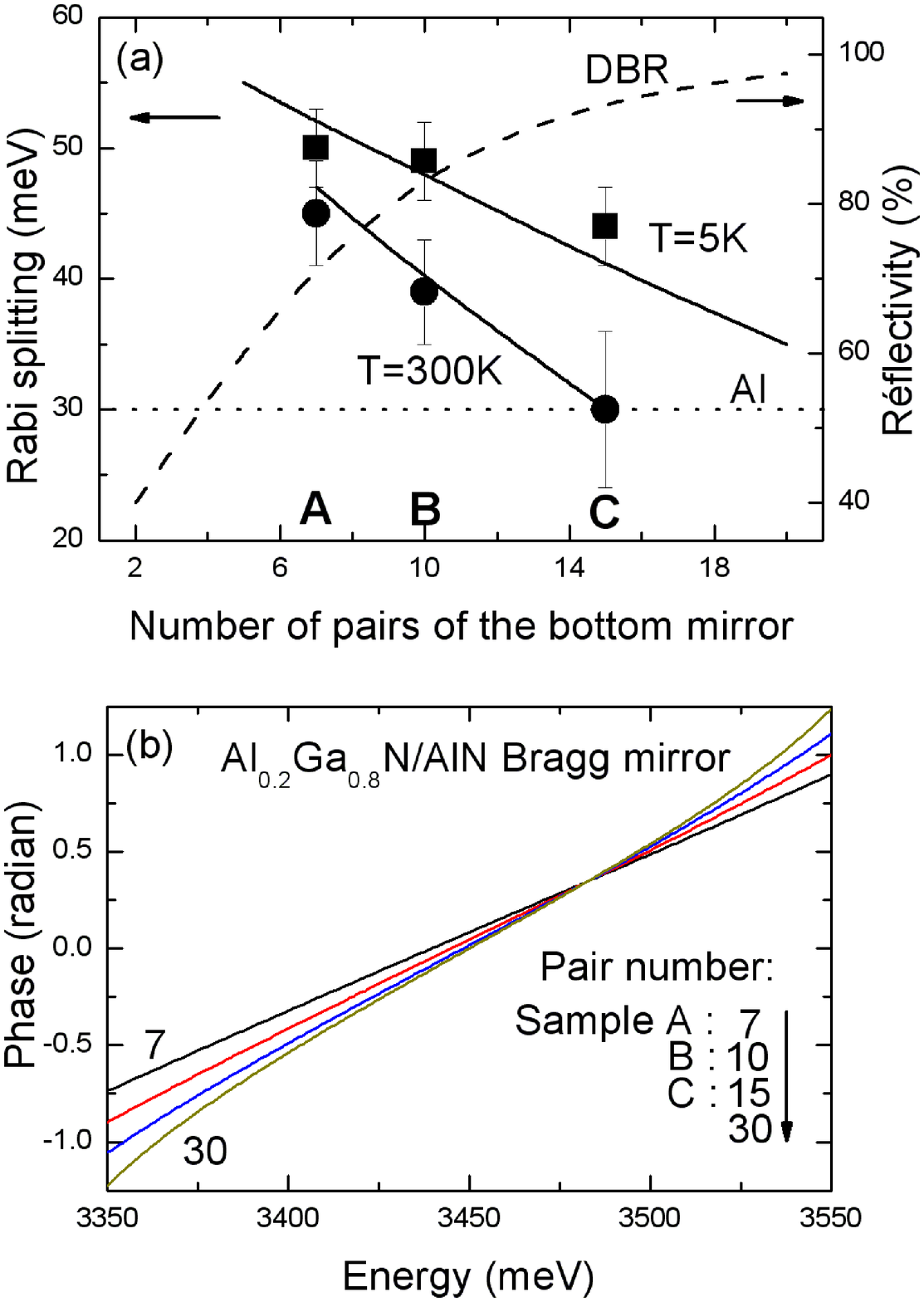}
				\caption{}
        \label{fig:Figure4}
\end{figure}

\noindent
The samples were also studied by reflectivity at room temperature. The experi\-mental and simulated spectra at 300 K in TE polarization for sample B are reported on figure~\ref{fig:SampleB}(c) and figure~\ref{fig:SampleB}(d) respectively. For the \hbox{simulation} of the room temperature reflectivity spectra, the same oscillator strength and inhomogeneous broadening parameters as in the 5 K case were used, while the homogeneous broadening parameter was increased to 15 meV, in agreement with its 300 K value in bulk GaN \cite{O.Aoudetobepublished}. At this temperature, the combination of the excitonic homogeneous and inhomogeneous broadening does not allow to separate the A and B excitonic dips any longer. 
The Rabi splitting deduced from the experiments is found to be 39$\pm$4 meV. 
The experimental spectra recorded in TE polarization for sample C whose bottom mirror contains 15 $\lambda$/2 pairs are reported on figures~\ref{fig:R_A771}(a) and \ref{fig:R_A771}(b) for low and room temperatures respectively. The Rabi splittings deduced are $44\pm2$~meV at 5 K and $30\pm5$~meV at 300 K. The large uncertainly of the latter is due to the difficulty to clearly distinguish the two modes at the resonance.
Figure~\ref{fig:Figure4}(a) reports the evolution of the Rabi splitting in samples A, B and C as a function of the number of pairs of the Al$_{0.2}$Ga$_{0.8}$N/AlN bottom mirror (the Rabi splitting of sample A has already been reported \cite{I.R.Sellers2006,ISBLLED}).
As shown on figure~\ref{fig:Figure4}(a), it decreases when the number of Al$_{0.2}$Ga$_{0.8}$N/AlN pairs increases, i.e. when the bottom mirror reflectivity increase. This result may appear counter-intuitive but it is nonetheless well accounted for by our transfer-matrix simulations at both 5~K and 300~K as  shown  by  the solid lines in figure~\ref{fig:Figure4}(a). It is important to recall that for each sample all the simulations are carried out with identical parameters (same cavity thickness and same excitonics parameters) except for the homogeneous broadening that changes with temperature. The reflectivity of the Al top mirror (R=52.5\%) is also reported as a dotted line on this figure together with the reflectivity of the bottom Bragg mirror as a function of the number of pairs (dashed line). In both cases, R is evaluated with respect to the GaN active layer and for an energy equal to 3.45~eV.
The cavity mode becomes narrower as the reflectivity of the number of $\lambda/2$ pairs increases; however, it corresponds to a decrease of the Rabi splitting. This can be qualitatively explained by the strong influence of the effective length of the cavity.
The latter is equal to : $L_{eff}=L_{DBR}+d$, where $d$ is the thickness of the active layer and $L_{DBR}$ represents the mirror penetration depth \cite{V.Savona1995,Panzarini}. At $\theta=0^{\circ}$, for an infinite pair number and a layer thickness equal to $\lambda/4$ the expression for $L_{DBR}$ is the following~:

\begin{align}
	L_{DBR}=\frac{1}{2}~\frac{\lambda_{0}}{n_{1}-n_{2}}
	\label{eqLdbr}
\end{align}

\noindent
This formula holds only for $n_{1}>n_{2}$ where $n_{1}$ and $n_{2}$ are respectively the refractive index of the Al$_{0.2}$Ga$_{0.8}$N and AlN layers.
For $\lambda_{0}=359$~nm, the mirror penetration depth for an infinite Al$_{0.2}$Ga$_{0.8}$N/AlN DBR calculated using Eq. (\ref{eqLdbr}) is equal to 795~nm. 
In order to estimate $L_{DBR}$ for samples A, B and C which have a finite numbers of mirror pairs, the phase of the DBRs was calculated in each case as a function of the photon energy within the tranfer-matrix model \cite{Panzarini}.
The results are shown on figure~\ref{fig:Figure4}(b) for 7, 10, 15 and 30 pairs bottom mirrors. In the energy range where the evolution of the phase is linear, the slope of the curve allows to determine $L_{DBR}$ \cite{V.Savona1995}. The $L_{DBR}$ value obtained for 30 pairs is 800~nm, in a good agreement with the value previously calculated for an infinite pair number (795~nm). For samples A, B and C the calculated mirror penetration depths deduced from figure 4(b) are respectively equal to 620~nm, 720~nm and 765~nm, i.e. it increases with the number of pairs. These values have to be compared with the thicknesses of the bottom mirrors~: 533~nm (A), 762~nm (B) and 1142~nm (C). Following the development given in reference~\cite{Vladimirova}, an approximate expression of the Rabi splitting as a function of the mirror penetration depth can be established as~:

\begin{align}
	\Omega\sim\sqrt{\frac{B~d}{\epsilon_{b}\left(d+\frac{L_{DBR}}{2}\right)}}
	\label{eqsplitting}
\end{align}

\noindent
where B is the oscillator strength and d the thickness of the active layer ($d\sim65$~nm in our $\lambda$/2 cavities).
This expression is calculated for an asymmetric cavity, i.e with a metallic top mirror and a DBR bottom one.
A quantitative comparison with the experiment is not possible because the previous formula holds only for high reflectivity mirrors ($1-R\ll 1$). However, it allows to qualitatively explain the experimental evolution of the Rabi splitting. The calculated ratio of Rabi splittings between A and C samples at 5K from equation (\ref{eqsplitting}) gives a value of 1.09 which compares favourably with the experimental ratio of 1.14.
Our transfer-matrix simulations account also well for this interpretation since the real structure of samples and both homogeneous and inhomogeneous excitonic broadenings are taken into account.\\
It is worth nothing that the evolution of $L_{DBR}$ as a function of the incidence angle ($\theta$) is opposite for TM and TE polarizations. The length $L_{DBR}$ increases for TM polarization with angle $\theta$ while it decreases for TE polarization \cite{Panzarini}. The experimentally determined Rabi splittings are found to be smaller in TM than in TE polarization for all the samples investigated in this work, in agreement with the angular dependence \cite{Panzarini}.\\



Near the GaN bandgap (3.45~eV), silicon can be considered as a poor metallic mirror with a reflection coefficient of $30\%$. The growth of only ten Al$_{0.2}$Ga$_{0.8}$N/AlN pairs on a silicon substrate allows to obtain a reflectivity ($R$) of $83\%$. If the same mirror were to be grown on a transparent substrate, for example an AlN substrate, sixteen Bragg pairs would be necessary to obtain $R=82\%$. Thus, the silicon substrate increases the efficiency of the lower Bragg mirror and a higher reflection coefficient can be achieved with a lower number of $\lambda/2$ pairs. Consequently, the reduction of the DBR thickness induces a smaller value of the mirror penetration depth.\\
\noindent
A solution to balance the cavity and obtain a better quality factor, could be to increase the thickness of the aluminium layer but the Al absorption is strong. The aluminium layer could however be used as a bottom mirror rather than a top one in order to decrease the cavity effective length and to obtain a larger Rabi splitting. However this way needs more technological steps for its achievement \cite{Duboz,rizzi:111112}.\\

\section{Optimization of the cavity quality factor}
\label{sec:OptimizationOfTheCavityQualityFactor}

\begin{figure}[!ht]
	\centering
			  \includegraphics[scale=0.83]{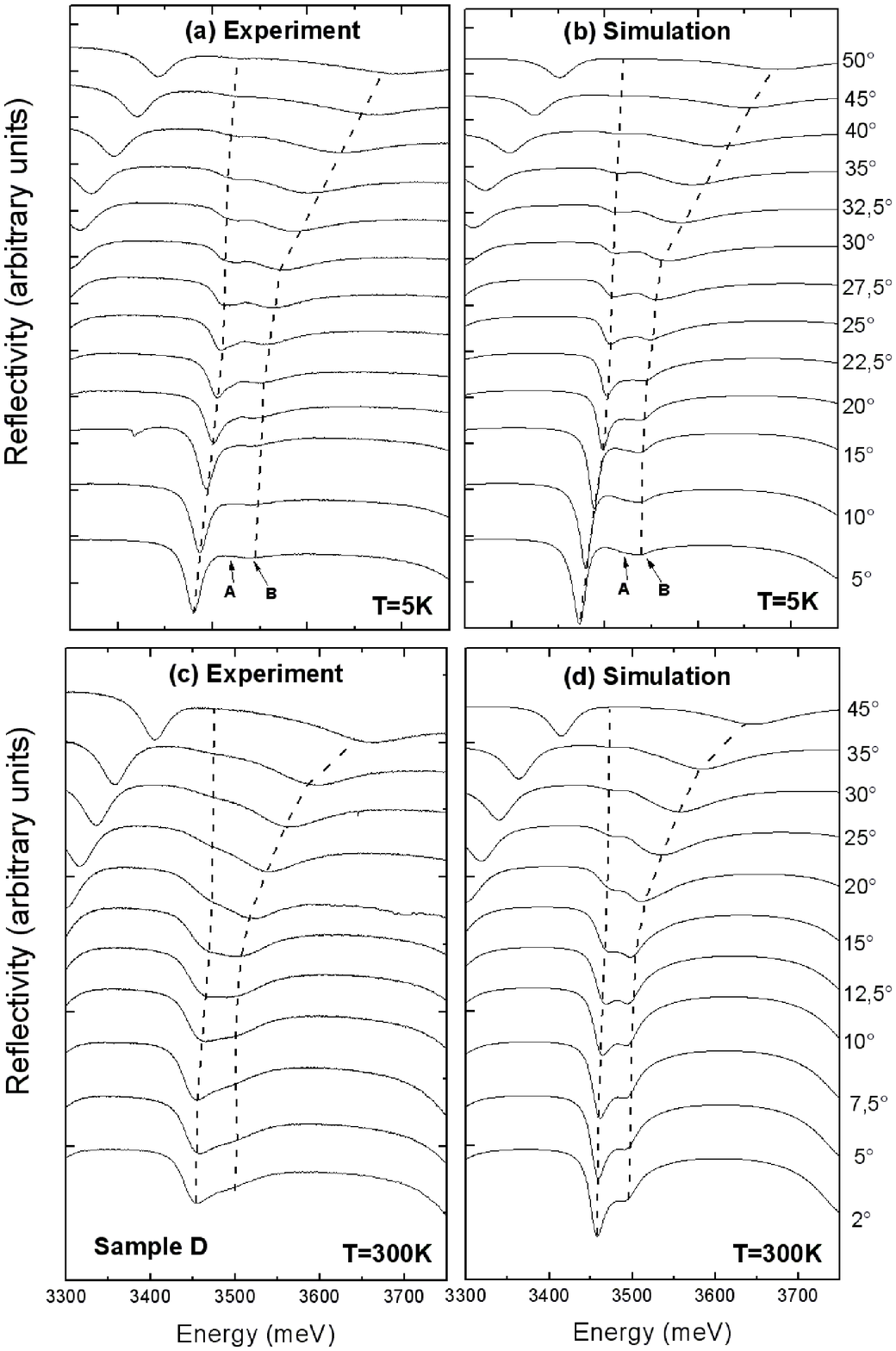}
				\caption{}		
				\label{fig:SampleD}
\end{figure}

The samples A, B and C studied in the previous section suffer from a low quality factor Q (varying from 50 in sample A to 60 in sample C) and this is detrimental for the observation of the strong coupling regime in emission, due to the lower coherence of the spontaneous emission compared to reflectivity \cite{I.R.Sellers}.\\
A fourth sample labelled D, with an improved Q, was elaborated from the same epitaxial stack as sample B by remplacing the Al top mirror by a dielectric mirror made of 8 $\lambda/2$ pairs of SiN/SiO$_{2}$. This structure is much more balanced because the reflectivity of the top mirror is equal to 90\% giving rise to a quality factor of 190. Angle-resolved reflectivity experiments have been performed at 5 K and at 300 K for both polarisations on this sample. Experimental spectra and simulations for TE polarisation are shown in figure \ref{fig:SampleD} (a and b) and (c and d) at 5 K and 300 K respectively. At 5 K, for the lowest incidence angle ($5^{\circ}$), the optical mode is slightly negatively detuned at $3480\pm1$~meV and much narrower than it is in the sample with an aluminium top mirror (sample B, Q = 55). The energies of A and B excitonic dips being identical in samples B and D, it can be concluded that the strain experienced by GaN is not modified by the deposition of the top dielectric mirror. 
The anticrossing between photons and excitons is clearly marked and occurs at $22.5^{\circ}$ with a Rabi splitting equal to $44\pm2$~meV. At 300 K, the strong coupling is also clearly in evidence and the Rabi splitting is equal to $28\pm5$~meV at $12.5^{\circ}$. 
The dielectric mirror increases the effective length of the cavity, and consequently the Rabi splitting is lower in sample D than it is in sample B. 
It is worth noting however that the ratio of the Rabi splitting to the lower polariton branch linewidth, i.e. the polariton visibility, is increased by a factor 3 at 5K in sample D in comparison with sample B (from 0.8 to 2.4).
We emphasize again that the simulations for samples B and D at both 5 K and 300 K are carried out with the same excitonic parameters (energy, oscillator strength, homogeneous and inhomogeneous broadenings).\\

\begin{figure}[!ht]
		\centering
        \includegraphics[scale=0.75]{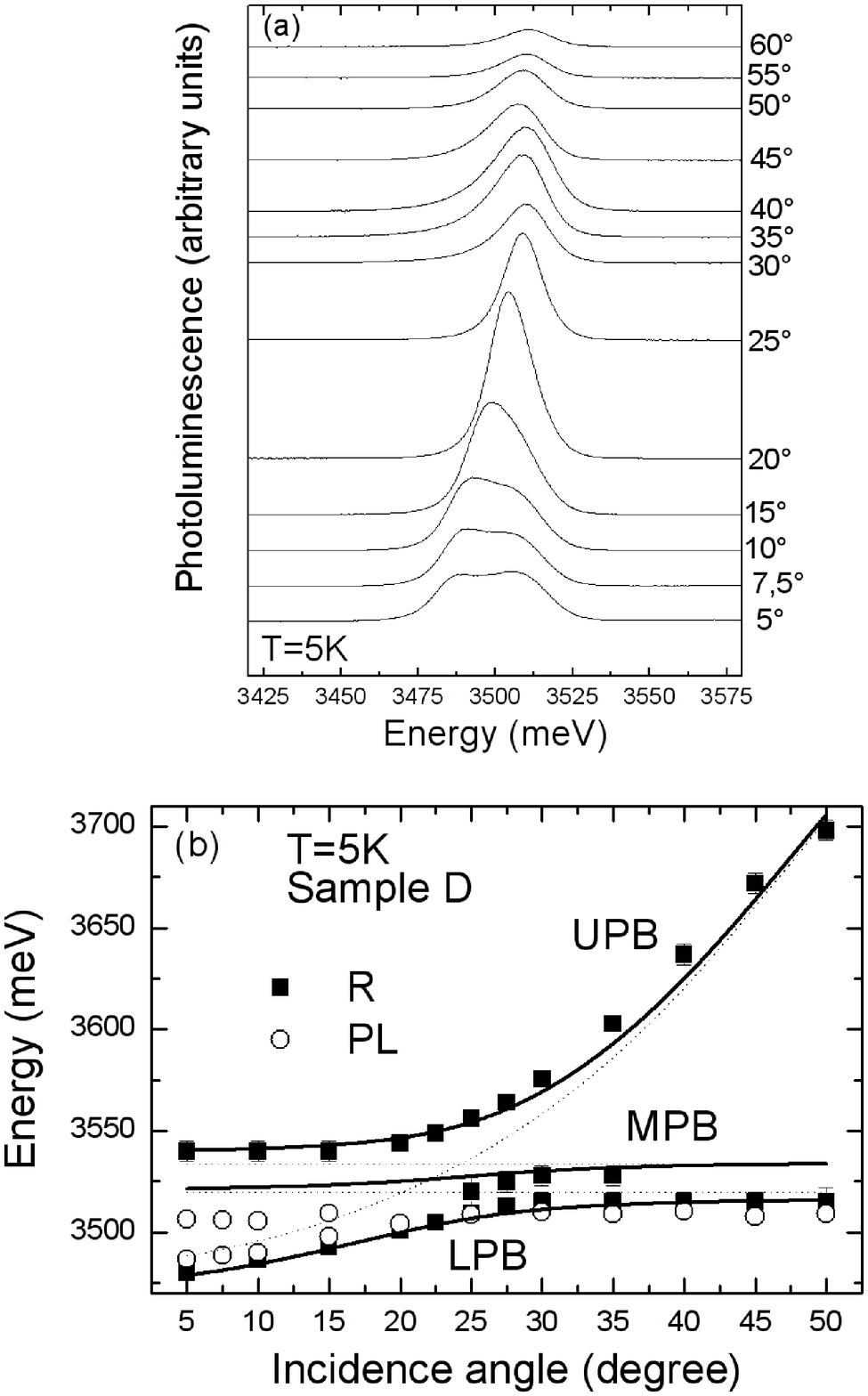}     
				\caption{}        
        \label{A808_diélec_Modèle_3x3}
\end{figure}

Angle-resolved photoluminescence (PL) experiments have been also performed at low temperature on samples B and D. Polariton emission has not been observed on sample B due to its low quality factor \cite{I.R.Sellers}. Only emission of localized uncoupled excitons has been observed whatever the incident angle \cite{ISBLLED}. The angle-resolved PL spectra of sample D are reported on figure \ref{A808_diélec_Modèle_3x3}(a). At low incident angle ($5^{\circ}$), two peaks are observed at $3487\pm2$~meV and $3506\pm2$~meV. The second peak disappears rapidly with increasing angle to the profit of the first one which can be followed from $5^{\circ}$ to $60^{\circ}$ degrees; it first blue-shifts with the increase of the incidence angle and then remains at the same energy for $\theta\geq 25^{\circ}$. The energy of the PL peaks together with the energy of the reflectivity dips at 5 K are displayed on figure \ref{A808_diélec_Modèle_3x3}(b) as a function of the incident angle. The dotted lines represent the energies of the uncoupled A and B excitonic and the photonic modes. The results of the 3x3 matrix diagonalisation are also reported as solid lines; the parameters used are $E_{ph}=3486$~meV at zero degree, $\Gamma_{ph}=18$~meV for the optical mode and $V_{A}=V_{B}=15$~meV for the interaction terms. The excitonic parameters are identical to those used for sample B. A good agreement is found between the reflectivity data and the calculations. The interaction terms deduced from the fit are lower than they are in sample B corresponding to a reduced Rabi splitting in sample D due to the larger effective length.\\
The low temperature spectrum of bulk GaN is often dominated by bound exciton (essentially donor-bound) recombinations \cite{leroux:3721}, so the second peak observed at 3506~meV is attributed to uncoupled localized excitons. It can be concluded that the strong coupling is also observed through PL experiments in sample D since the evolution of the first PL peak as a function of incident angle coincides with the evolution of the lower polariton branch (LPB) observed by reflectivity. However, because of the low temperature, and also the low excitation intensity ($\sim~6$~W/cm$^{2}$), only the lower polariton branch can be detected in photoluminescence.\\

\section{Conclusion}
\label{sec:Conclusion}

The influence of the mirror-type and reflectivity on the strong light-matter coupling regime in $\lambda/2$ bulk GaN microcavities has  been  studied. In the case of structures with a semi-transparent aluminium top mirror, the increase of the number of $\lambda/2$ (Al,Ga)N/AlN pairs of the bottom mirror is correlated with a decrease of the Rabi splitting. This result is qualitatively explained by an increase of the ligth penetration depth of the bottom mirror. The transfer-matrix formalism, which allows to consider the real structure of the samples accounts well for all the reflectivity spectra and supports our interpretation. An Al layer as a top mirror reduces the cavity length and allows a large Rabi splitting but does not give an optical quality factor sufficient for the observation of the polaritonic emission. However, replacing the metallic top mirror by a dielectric DBR allows to increase the quality factor to 190. The Rabi splitting at low temperature is slightly reduced but the emission of the lower polariton branch can be observed.

\begin{acknowledgments}
The authors are grateful to J.-Y. Duboz for useful discussions and suggestions, and acknowledge H. Ouerdane and G. Malpuech for a critical reading of the manuscript.
\end{acknowledgments}

\newpage

\noindent
\textbf{\large{Figure captions:}}\\

\noindent
\textbf{Fig. 1:} Angle-resolved reflectivity spectra recorded in TE polarization and simulations within the transfer-matrix formalism for samples B at low temperature (a,b) and room temperature (c,d). The dashed lines are only guides for the eye allowing to follow the lower and upper polariton branches.

\noindent
\textbf{Fig. 2:} Experimental energy of the optical and excitonic modes as function of the incidence angle at 5 K in TE polarization for sample B. The solid lines represent the eigenvalues of the 3x3 matrix simulating the exciton-photon coupling. The dotted lines are related to the uncoupled optical and excitonic modes.

\noindent
\textbf{Fig. 3:} Angle-resolved reflectivity spectra recorded in TE polarization for sample C at low temperature (a) and room temperature (b). The dotted lines are only guides for the eye allowing to follow the lower and the upper polariton branches.

\noindent
\textbf{Fig. 4:} (a) Experimental Rabi splittings at 5 K and 300 K for samples A, B and C as a function of the number of pairs of the bottom mirror. The solid line corresponds to calculations within the transfer matrix formalism. Reflection coefficients of the top (dotted line) and bottom (dashed line) mirrors are also shown. (b) Phase of the reflection coefficient of the bottom mirror as a function of photon energy for various $lambda/2$ Al$_{0.2}$Ga$_{0.8}$N/AlN number of pairs (7, 10, 15 and 30).

\noindent
\textbf{Fig. 5:} Angle-resolved reflectivity spectra recorded in TE polarization and simulations within the transfer-matrix formalism for sample D at low temperature (a,b) and room temperature (c,d). The dashed lines are only guides for the eye allowing to follow the lower and the upper polariton branches.

\noindent
\textbf{Fig. 6:} (a) Low temperature (5 K) photoluminescence spectra of sample D as function of the emission angle. (b) Energies of the optical and excitonic modes deduced from the 5 K reflectivity (solid squares) and the photoluminescence (empty circles) experiment on sample D as a function of the angle. The solid lines represent the eigenvalues of the 3x3 matrix used for simulating the exciton-photon coupling. The dotted lines are related to the uncoupled optical and excitonic modes.

\newpage

\end{document}